\documentclass[prl,aps,twocolumn,preprintnumbers,showpacs,superscriptaddress]{revtex4}
\usepackage[colorlinks=true, pdfstartview=FitV, linkcolor=red, citecolor=blue, urlcolor=blue]{hyperref}

\usepackage{graphics,graphicx,epsfig,bm}

\usepackage{epstopdf}

\begin{document}

\title{Independent Noise Approximation for Spin-Boson Decoherence}

\date{\today}

\author{Man-Hong Yung}
\email[email: ]{myung2@uiuc.edu}

\affiliation{Department of Physics, University of Illinois at
Urbana-Champaign, Urbana IL 61801-3080, USA}


\begin{abstract}
Quantum error correction is a solution to preserve the fidelity of quantum information encoded in physical systems subject to noise.  However, unfavorable correlated errors could be induced even for non-interacting qubits through the environment (bath),  when they are ``packed" together. The question is, to what extent can we treat the noise induced by the bath as independent? In the context of the spin-boson model, we show that, under some reasonable constraints, the independent noise approximation could be valid. On the other hand, in the strongly correlated limit, we show how the method of decoherence free subspace can be made applicable. Combining these two methods makes fault-tolerant quantum computation promising in fighting against correlated errors.   

\end{abstract}

\maketitle
Quantum information encoded in realistic physical systems is very fragile, compared with the classical counterparts. Quantum error correction \cite{Shor95,Steane96} is possible and efficient provided that the errors are independent or weakly correlated.  For independent noise, where errors are modeled as stochastic events, error thresholds \cite{Knill98, Aliferis05, Aharonov06} can be estimated to give an upper bound of the error rate for fault-tolerant quantum computation. The effects of correlated errors are generally not favorable for error correction as more restrictive error thresholds are needed. In reality, quantum noise may not behave like stochastic noise --- the independent noise assumption is questionable. No physical qubit can be genuinely isolated from the environment (bath).  In particular, for solid state systems, if the qubits are interacting with the same bath, indirect correlations could be generated, and the protection from error correction codes becomes ineffective. Our goal here is to illustrate the essential features and general structures of the environment induced decoherence for the purpose of quantum error correction.

This work is motivated by the recent studies \cite{Aliferis05, Aharonov06, Terhal05, Klesse05, Novais06} related to this problem. However, some physical considerations have been ignored, and consequently the adverse effects of the bath on quantum error correction were often {\it overestimated}. In \cite{Aliferis05, Aharonov06, Terhal05}, attempts of deriving mathematically rigorous error thresholds were made by summing over the so-called ``fault paths". However, for general environments, the spin-bath coupling $\left\| {H_{SB} } \right\|$ is unbounded and hence fails the analyses. The spin-boson model \cite{Leggett87} (with [$H_S,H_{SB}] =0$ cf. Eq. (\ref{H})) have been considered in \cite{Klesse05, Novais06}, where a coarse-grained description on the dephasing effects were studied in \cite{Klesse05}, but the possibility of reducing the collective effects (also ignored in \cite{Novais06}) through decoherence free encoding were not considered. Moreover, the effects of including non-trivial local Hamiltonians $H_S$ have not been properly explored, as we shall see it does play an important role in generating spatial error correlation. This work is aimed at providing a more complete and more physical picture on this problem, and suggests a more optimistic view on the effects of environment induced decoherence. Particularly, we shall address two questions: (1) To what extent, can the noise induced by the bath be considered as independent? (2) For fully correlated noise, how do we minimize the decoherence effects? 

We formulate the problem in terms of the relevant two-point correlation functions. For a generic boson bath (e.g. acoustic phonon bath) and physical qubits with local energy splittings $\Delta$, we found that the error correlation generated from the energy-converving processes (here called ``bit-flipping" case) becomes significant when the qubit-qubit separations are less than a length scale $\lambda_* \equiv \hbar c / \Delta$, where $c$ is the wave speed of the bath. The physical origin of the spatial error correlation is due to the {\it constructive interference} of the local disturbances from the qubits to the bath. This provides an answer to question (1). For question (2), we found that when $\Delta \to 0$ (here called dephasing case), the errors are fully correlated when the long wavelength modes of the bath become dominant. Provided that the effective qubit-qubit interactions [cf. Eq.(\ref{H_eff})] are properly handled, it is possible to minimize the effects by using the noiseless subspace (or decoherence free subspace DFS) to encode the qubits.    

\emph{Definition of Independent Noise ---} We first clarify the concept of error correlation in the context of quantum error correction. The degree of error correlation may be best quantified by the deviation from the independent noise. We motivate the definition of `independent noise' through a simple example: consider a register of qubits, consisting of $N$ spin-1/2 particles, initialized in some encoded state $\left| {Qubits} \right\rangle$, subject to a random classical magnetic field which causes an unwanted evolution
\begin{equation}\label{U}
U = \prod\limits_{j = 1}^N {\exp \left( {\epsilon _j W_j } \right)}  \approx I + \epsilon _j W_j  + \epsilon _j \epsilon _k W_j W_k  + ...
\end{equation} 
where $W_k = \{ X_k,Y_k,Z_k\}$ is one of the Pauli matrices ($W_k^2=I$) acting on spin $k$ and the random variable $\left| {\epsilon _k } \right| \ll 1$ is assumed to be small. In powers of $\epsilon_k$, each term in this series is in general not mutually orthogonal. However, after error syndrome detection, the final state will be projected into some states with definite parity with respect to the corresponding stabilizers. The probability $P_1(W_k)$ of single error on spin $k$  can be estimated by the amplitude square $A_1(W_k)^2  \equiv \left\| {\epsilon _k W_k \left| {Qubits} \right\rangle } \right\|^2  = \left| {\epsilon _k } \right|^2 
$ of the $W_k$ term. The amplitude of any two-qubit error is  simply a product of the amplitudes of two single-qubit errors $A_2\left( {W_j W_k } \right) = | \epsilon_j \epsilon_k | =  A_1\left( {W_j } \right)  A_1\left( {W_k } \right) $, for any $j \ne k$. Provided that all $\left| {\epsilon _k } \right| \ll 1$, the standard threshold theorem about independent stochastic noise should be valid. To generalize the above argument to the errors caused by the noise from the environment (formally denoted as $\left| {Env} \right\rangle$), we may (naively) promote  the complex numbers $\epsilon_k$ to operators $\hat \epsilon_k$ acting on the environment. Then, $A_1(W_k)^2  = \left\langle {Env} \right|\hat \epsilon _k^\dagger \hat  \epsilon _k \left| {Env} \right\rangle $ and so on. However, generally  $A_2 \left( {W_j W_k } \right) \ne A_1 \left( {W_j } \right)A_1 \left( {W_k } \right)$, but we want to know under what circumstances the equality would hold. This motivates our definition of the independent noise model:
\begin{equation}\label{A}
A_n \left( {W_1 W_2 ... W_n} \right) \approx A_1 ( {W_1 }) A_1 ( {W_2 } ) \cdots A_1( {W_n }) \quad 
\end{equation}
for all possible combinations of the Pauli matrices. The `$\approx$' sign offers us the flexibility to neglect higher order corrections, if any. Below we shall study the question about the validity of the independent noise approximation in the context of the spin-boson model.  The method to be introduced may also be applicable to more general models of the environment, and shall likely give the same qualitative results.  

\emph{Spin-Boson Model ---} We assume that the interaction between the boson bath and the physical qubits is sufficiently ``weak"  to be considered as perturbation. The full Hamiltonian $H=H_S+H_B+H_{SB}$, or
\begin{equation}\label{H}
H = H_S  + \sum\limits_k {\hbar \omega _k a_k^\dagger  a_k }  + g\sum\limits_k {\left( {{\tilde Z}_k a_k^ \dagger   + {\tilde Z}_k^\dagger  a_k } \right)}
\end{equation} 
 consists of three parts, respectively, $H_S$ (to be specified later) the system Hamiltonian of $N$ physical qubits, $H_B $ the bosonic bath Hamiltonian, and $H_{SB} $ the interaction Hamiltonian, ${\tilde Z}_k  \equiv \sum\nolimits_{j = 1}^N {Z_j  e^{ {-i\vec k \cdot \vec r_j }}}$, where $\vec r_j$ is the position of the $j$ spin. For simplicity  we assume that all of the spins are coupled uniformly $g_j=g$ with the bath. Morevoer, we shall confine our attention to the baths having two generic properties: (a) the modes are 3-dimensional $\vec k = (k_x,k_y,k_z)$ and (b) the modes have a linear dispersion relation $\omega_k = c k$.

\emph{Case I: Dephasing ---} To continue with a more quantitative analysis, here we shall first consider the $H_S = 0$ case, since it is exactly solvable \cite{Palma96, Reina02} (alternatively, take the $\Delta \to 0$ limit from Eq. (\ref{S})) and contains rich physical  contents. The unitary operator ($\hbar=1$) $U=\exp(- i H t)$ can be decomposed into three parts
\begin{equation}\label{3_parts}
U\left( t \right) =  \exp({ - iH_B t}) \exp{iH_{eff} (t)} \exp{G_{SB}(t)}  \quad,
\end{equation}
where  $H_{B}$ is defined in Eq. (\ref{H}), 
\begin{equation} 
G_{SB} \left( t \right) = \sum_{j = 1}^N {Z_j \phi _j \left( t \right)} \quad,
\end{equation} 
and with $f_k\left( {r_j ,t} \right) \equiv (g/\omega_k) e^{ - i\vec k \cdot \vec r_j } \left( {1 - e^{i\omega _k t} } \right)$, 
\begin{equation} 
\phi _j \left( t \right) \equiv \sum\limits_k {\left[ {f_k \left( {r_j ,t} \right)a_k^ \dagger   - f_k \left( {r_j ,t} \right)^* a_k } \right]}  \quad .
\end{equation}
The effective spin-spin interaction is of the $ZZ$ form:
\begin{equation}\label{H_eff}
H_{eff}  = \sum\limits_k {\frac{{g^2 }}{{\omega _k^2 }} {\tilde Z}_k {\tilde Z}_k^ \dagger   \left[ {\omega _k t - \sin \left( {\omega _k t} \right)} \right]}  \quad.
 \end{equation}
 It is generated through the virtual transitions of the spin-bath interaction. By definition, it does not depend on the state of the bath and the effects are predictable, and hence should not be considered as noise \cite{Leggett_noise}. In principle, it can be included as part of the quantum circuit, or eliminated through some refocusing schemes well-developed in NMR quantum computing. For the moment, we shall neglect it. The term $\exp(-i H_B t)$, containing no spin variables, is irrelevant for evaluating the error amplitudes and will also be neglected. Our goal is to expand  the remaining parts in a power series, similar to Eq. (\ref{U}), of $g$ (or $Z_j$), and then estimated the relative size of each term. 

We assume the modes are isotropic, i.e., $\omega_{-k}=\omega_k$. Then all commutators  $\left[ {\phi _j ( t),\phi _k ( t)} \right]=0$, containing odd parity terms $\sin [ {\vec k \cdot \left( {\vec r_j  - \vec r_k } \right)} ]$, should vanish. This means
\begin{equation}\label{G}
\exp{G_{SB} ( t)}  = \prod\limits_{j = 1}^N {\exp [{Z_j \phi _j ( t )} ]}  \quad,
\end{equation}
which resembles Eq. (\ref{U}). Suppose at $t=0$ the spins are in a pure state and the bath is in a thermal state. The square of the error amplitudes are given by  ($\phi^\dagger = - \phi$)
\begin{equation}
A_n \left( {Z_1 Z_2 ...Z_n } \right)^2  = \left\langle {\phi _1 ^\dagger \phi _1 \phi _2 ^\dagger \phi _2 ...\phi _n ^\dagger \phi _n} \right\rangle \quad . 
\end{equation}
From the Bloch identity \cite{Mermin66}, we have $\left\langle {\exp \xi _j \phi _j } \right\rangle  = \exp \left[ \left( { \xi _j \xi _m /2} \right)\left\langle {\phi _j \phi _m } \right\rangle \right]$. Here repeated indices imply summation, and $\xi_j$ is an arbitrary complex number. The quantity $A_n^2$ can then be evaluated by the standard technique of summing different ``pairing" of the two-point correction functions, which are the key quantities to be considered for error correlation, 
\begin{equation}\label{2-point}
\left\langle {\phi _j \phi _m } \right\rangle  = \sum\limits_k {\left[ {h_k( {R,t})\langle {a_k^ \dagger  a_k }\rangle  + h_k( {R,t})^* \langle {a_k a{_k^ \dagger} } \rangle } \right]} ,
\end{equation}
where $h\left( {R,t} \right) = (2 g/\omega_k)^2  e^ { - i\vec k \cdot \vec R}\sin ^2 \left( {{\textstyle{1 \over 2}}\omega _k t} \right)$ and $\vec R \equiv {\vec r_j  - \vec r_m}$. If for all $j \ne m$, $\left\langle {\phi _j \phi _m } \right\rangle =0$, then we recover the independent noise condition Eq. (\ref{A}). However, this is not generally true. Although the phase factor  $e^ { - i\vec k \cdot \vec R}$ tends to cause lots of cancellations, the remaining term tends to suppress this effect. Recall that $\sin ^2 \left( {\omega _k t/2} \right)/\omega _k^2  \to \left( {\pi /2} \right)\delta \left( \omega_k  \right)t$, for $t \gg t_{\max} \equiv \max \{\omega_c^{-1}, R/c, \hbar/k_B T \}$ being the longest time scale. Roughly speaking, constructive interference occurs (i.e., $| \left\langle {\phi _j \phi _m } \right\rangle | \sim | \left\langle {\phi _j \phi _j } \right\rangle | $, $j\ne m $) whenever the waves of disturbance has enough time to travel from spin $j$ to spin $m$, i.e., $t \gtrsim t_s^{\max}\equiv \max \{\left| {\vec r_j  - \vec r_m } \right|/c\}$. In the case of phonon bath, if the spins are located in a region of area $100 \mu m^2$,  then $t_s^{\max} \approx 10^{-7} s$, which suggests that to maintain the independent noise approximation, the {\it whole} error correction procedure has to be executed at least $10^7$ times per second.

From the point of view of current experimental situation, it is not unreasonable to consider the ``worst case" scenario where we set 
\begin{equation}\label{j=m}
\left\langle {\phi _j \phi _m } \right\rangle  = \left\langle {\phi _j \phi _j } \right\rangle = \left\langle {\phi _m \phi _m } \right\rangle 
\end{equation}
for all $j$ and $m$, and consider the impacts to quantum error correction. In this case, the probability  (upper bound)  $P_n \le A_n^2$ of $n$-qubit errors would be enhanced (relative to the independent case $P_n = P_1^n$) by a factor $\left( {2n} \right)!/2^n n! \approx \sqrt 2 \left( {2n/e} \right)^n $ for large $n$ (e.g. $n=2^k$ for $k$-level concatenated distance-$3$ codes). Consider the standard error threshold analysis based on the independent noise model. The failure probability is given by $P_{fail} = P_{th} (P_1/P_{th})^n$, where $P_{th}$ is a threshold error rate, which depends on the coding methods and circuit designs. When $P_1<P_{th}$, $P_{fail}$ can be made arbitrarily small by increasing the levels of concatenation. Now in the presence of the constructive interference effects, $P_1^n  \to \sqrt 2 \left( {2n/e} \right)^n P_1^n$, hence
\begin{equation}\label{P_f}
P_{fail}  \approx \sqrt 2 P_{th} \left( {\frac{{2n}}{e}\frac{{P_1 }}{{P_{th} }}} \right)^n \quad.
\end{equation}
The concatenation method becomes inefficient when $P_1  \lesssim \left( {e/2n} \right)P_{th} $ and {\it breaks down beyond that}. Therefore, the interference effect imposes a more stringent  threshold for quantum error correction. 

The good news is that, in the same limit, provided that the effective interaction [cf. Eq. (\ref{H_eff})] is properly compensated, the correlated errors can be made vanished within the decoherence free subspace (DFS) \cite{Palma96,Lidar98}. In the ordinary analysis of DFS, one requires $H_{SB} \left| {\rm DFS} \right\rangle  = U\left| {\rm DFS} \right\rangle$, where $U$ is either zero or some unitary operators acting only on the qubits. This condition, requiring zero qubit-qubit separation, is not assumed in our case. However, the qubits do effectively ``see" the same environment. Since when the long wavelength modes $k \to 0$  become dominant, the spatial qubit separations cannot be ``resolved". To justify this, we consider the subspace of states satisfying $\sum\nolimits_{j=1}^N {Z_j \left| {DFS} \right\rangle }  = 0$, e.g. all symmetrized states with equal numbers of `0' and `1'. From Eq. (\ref{j=m}), $\left\langle {G_{SB}^2 } \right\rangle_{DFS}  = 0$ and hence by invoking the Bloch identity again
\begin{equation}
\left \langle {\exp{G_{SB}} } \right \rangle _{DFS}  \, = \, I  \quad.
\end{equation}
It is now clear from Eq. (\ref{U}) that the spin-bath dynamics is decoupled i.e., $U\left( t \right)\left| {DFS} \right\rangle  \otimes \left| {Env} \right\rangle  = e^{ iH_{eff} \left( t \right)} \left| {DFS} \right\rangle  \otimes e^{ - iH_B t} \left| {Env} \right\rangle$. Therefore, the idea of isolating the effective interaction gives us new insights about the method of DFS (answer to question (2)) and makes it physically more applicable.

\emph{Case II: Bit-Flipping --- } In the following, we shall consider the effects of including $H_S \ne 0$. It is obvious that we reach the same conclusions if $[H_S,H_{SB}]=0$. The simplest non-trivial case seems to be $H_S  = \sum\nolimits_{j = 1}^N {(\Delta/2) X_j }$. As mentioned before, the coupling with the bath causes transitions between the two eigenstates of $X$, and is most efficient for $\hbar \omega_k \approx  \Delta$, as $\Delta$ sets the energy and hence the distance scale $\lambda_* = \hbar c / \Delta$ for our problem. For weak spin-bath coupling $H_{SB}$, we employ the standard method of canonical transformation: $ {\tilde H} \equiv e^{-S} He^{  S}  = H_S+H_B  + \left( {1/2} \right)\left[ {H_{SB} ,S} \right] + ...$, where $S$ satisfies the relation $H_{SB}+[H_S+H_B,S]=0$. Define $L_j^ \pm   \equiv \left|  \pm  \right\rangle \left\langle  \mp  \right| = \left( {Z_j  \mp iY_j } \right)/2$  and $S \equiv \sum\nolimits_{j = 1}^N {S_j } $, we write ($\hbar =1$)
\begin{equation}\label{S}
S_j  = g\sum\limits_k {\left[ {T_j \left( {\omega _k } \right)e^{ - i\vec k \cdot \vec r_j} a_k^\dagger   + T_j \left( { - \omega _k } \right)e^{i\vec k \cdot \vec r_j} a_k } \right] ,}  
\end{equation}
where $T_j \left( {\omega _k } \right) \equiv L_j^ -  /\left( {\Delta  -  \omega _k } \right) - L_j^ +  /\left( {\Delta  +  \omega _k } \right)$. The evolution operator can be expressed as $U\left( t \right) = \exp S\exp ( { - i\tilde Ht} )\exp \left( { - S} \right)$. The (lowest order) effective interaction can be extracted from the commutator $\left[ {H_{SB} ,S} \right]$, it turns out to be of the $ZZ$ form:
\begin{equation}\label{H_jm}
H_{jm}  = Z_j Z_m \sum\limits_k {2g^2 \frac{{ \omega _k }}{{  \omega _k ^2  - \Delta ^2 }}} \cos \left( {\vec k \cdot \vec R} \right) \quad ,
\end{equation}
which reduces to the first term of Eq. (\ref{H_eff}) when $\Delta \to 0$. The second term, containing $\sin(\omega_k t)$, emerges when we combine and rearrange the operators in $U$ and transform it into the form (with $H_B  \to H_B  + H_S $) in Eq. ({\ref{3_parts}}). Crucially, since all diagonal matrix elements $\left\langle n \right|H_{jm} \left| n \right\rangle  = 0$ are zero in the eigenbasis of $H_S$, the term $H_{jm}$ does not grow linearly in time (bounded by a factor $1/\Delta$) and becomes negligible \cite{Order} for $\Delta t / \hbar \gg 1$. 

We shall now consider the interference effects. For the moment, we neglect the higher order corrections and any effective interaction generated, and write $U(t) \approx e^{ - i(H_S  + H_B )t} e^{ S\left( t \right)} e^{- S\left( 0 \right)} $, where $S\left( t \right) = e^{i(H_S  + H_B )t} S\left( 0 \right)e^{ - i(H_S  + H_B )t} $. We are mostly interested in the limit where the energy non-conserving terms in Eq. (\ref{S}) becomes relatively small. Then, the counterpart of Eq. (\ref{G}) is
\begin{equation}
\exp G_{SB} \left( t \right) = \prod\limits_{j = 1}^N {\exp \left[ {i Y_j \varphi _j^y \left( t \right)} \right]} \exp \left[ {Z_j \varphi _j^z \left( t \right)} \right] ,
\end{equation}
where, with $\eta =0$ for $\varphi_j^z$ and $\eta =1$ for $\varphi_j^y$,
\begin{equation}
\varphi _j^{z,y} \left( t \right) \equiv \sum\limits_k {\left[ {\tilde f_k \left( {r_j ,t} \right)a_k^\dagger   - \left( { - 1} \right)^\eta  \tilde f_k \left( {r_j ,t} \right)^* a_k } \right]} ,
\end{equation}
and $\tilde f_k \left( {r_j ,t} \right) \equiv [g/(\omega _k  - \Delta )]e^{ - i\vec k \cdot \vec r_j } [1 - e^{i(\omega _k  - \Delta )t} ]$. Now the same argument, about the error upper bound, following Eq. (\ref{G}) should go through the same way. However, the two-point correlation functions $\left\langle {\varphi _j^{z,y} \varphi _m^{z,y} } \right\rangle$, which cause spatial error correlation, vanishes when the qubits are separated sufficiently far apart $\Delta t_s / \hbar \gg 1$, where $t_s  \equiv \left| {\vec r_j  - \vec r_m } \right|/c$. For example, $\left\langle {\varphi _j^z \varphi _m^z } \right\rangle$ is almost the same as in Eq.(\ref{2-point}), except the replacement $h_k \left( {R,t} \right) \to \tilde h_k \left( {R,t} \right)$ where
\begin{equation}
\tilde h_k \left( {R,t} \right) \equiv \frac{{4g^2 }}{{\left( {\omega _k  - \Delta } \right)^2 }}e^{ - i\vec k \cdot \vec R} \sin ^2 \left[ {{\textstyle{1 \over 2}}\left( {\omega _k  - \Delta } \right)t} \right] .
\end{equation}
When $t \gg t_{\max} \equiv \max \{\omega_c^{-1}, t_s, \hbar/k_B T\}$, we have $\left\langle {\varphi _j^z \varphi _m^z } \right\rangle  \approx \left\langle {\varphi _j^z \varphi _j^z } \right\rangle \sin \left( {\Delta t_s } \right)/\Delta t_s $. In fact, to have the interference terms get cancelled, the condition $\Delta t_s /\hbar \gg 1$ could be too strong. It is likely that for $\Delta t_s /\hbar \sim \pi$, the correlation functions start to change signs and cause lots of cancelations. This however should depend on the spatial distribution of the physical qubits. Under these conditions, we expect that the independent noise approximation is valid (answer to question (1)) for the spin-boson decoherence, provided that the higher order terms, so far neglected, are also small compared with that generated by the single-error terms discussed [cf. Eq. (\ref{S})] above. Below we shall give a heuristic argument to justify that. 

In the canonical transformation, higher order corrections terms are generated from the series of commutators $\left[ {\left[ {\left[ {H_{SB},S} \right],S} \right],...,S} \right]$. For each term, we denote $(n,m)$ to represent the number $n$ of Pauli matrices (or `weight'), and $m$ the weight of $a^\dagger$ and $a$, for example  $(1,1)$ for both $S$ and $H_{SB}$, which are $O(g)$. Taking the commutator with $S$, each term changes from $(n,m)$ to either $(n,m+1)$ or $(n + \xi,m-1)$ where $\xi=\{0,\pm 1\}$, corresponding to taking the commutator for the Pauli matrices and the bath operators respectively. The case $(n+1,m-1)$ is the only case causing an increase in the weight of the Pauli matrices, with a cost of increasing one power in $g$ (making it smaller). The effective interaction $(2,0)$ in Eq. (\ref{H_jm}) is an example of this. Starting from $H_{SB}$, i.e., (1,1), apart from $(2,0)$, any $(n>1,m)$ term is at least $O(g^{n+1})$, which is smaller than the contribution obtained by $S^n$ which is $O(g^n)$. We therefore conclude that the higher order corrections are negligible for sufficiently small $g$. 

Lastly, we remark that direct qubit-qubit interaction can also be included in $H_S$ following the same line of thought, and the results should be qualitatively the same, except (not surprisingly) that the errors on the two interacting qubits become strongly correlated. On the other hand, it can also be generalized to determine the non-Markovian behaviors of the bath between the error-correcting cycles. In this case, the space-time correlation functions, $\left\langle {\varphi _j \left( {t_a } \right)\varphi _m \left( {t_b } \right)} \right\rangle$ where $t_a\ne t_b$, are needed to be considered. Details to be given elsewhere.   

\emph{Conclusions ---} In conclusion, we have illustrated the general structure and essential features of the spin-boson decoherence for quantum error correction. The noise correlations are determined by the relevant two-point correlation functions. In particular, we have considered two special cases where the correlated errors can either be generated freely in the dephasing case or be suppressed completely (when $\Delta t_s /\hbar \gg 1$) in the bit-flipping case. In the former case the error threshold becomes much more stringent [cf. Eq. (\ref{P_f})] , but it could be handled with a modified decoherence free subspace (DFS) method. In reality, environment induced decoherence may exhibit features of both cases. It is likely that the combination \cite{combination} of the two methods, namely strong local field (hardware) and error-correcting codes together with DFS (software), could help to minimize more general errors. Based on the results of this work, spatially correlated errors do not seem to be a fundamental barrier for fault-tolerant quantum computation.



\begin{acknowledgments}
M.H.Y acknowledges the support of the NSF grant EIA-01-21568 and the Croucher Foundation, and thanks  R. Laflamme for the hospitality of the Institute for Quantum Computing where part of this work is done. M.H.Y also thanks J. Emerson, D. Gottesman,  J. Preskill, D. Leung, N. Shah and F. Wilhelm for valuable discussions, and especially A. J. Leggett for comments and criticisms.
\end{acknowledgments}




\begin{thebibliography}{99}
\bibitem{Order}
At $T=0$, the single-qubit error rate is $\sim O(J(\Delta) t)$ while $H_{eff}$ is $\sim O(J(\Delta)/\Delta^2 t_s)$.
\bibitem{combination}
For example, symmetrizing all state $\left| {0} \right\rangle  \to \left| {01} \right\rangle$ and  $\left| {1} \right\rangle  \to \left| {10} \right\rangle$, will fall into the DFS w.r.t. the collective $Z$-errors. The potential  collective $X$ and $Y$ errors could be suppressed by the local potentials $\Delta Z$.
\end{thebibliography}
\end{document}